\documentclass{ws-ijmpa}
\usepackage[super,compress]{cite}
\usepackage{graphicx}

\def\bea{\begin{eqnarray}}
\def\eea{\end{eqnarray}}
\def\beq{\begin{equation}}
\def\eeq{\end{equation}}
\def\bq{\begin{quote}}
\def\eq{\end{quote}}
\def\gappeq{\mathrel{\rlap {\raise.5ex\hbox{$>$}} {\lower.5ex\hbox{$\sim$}}}}
\def\lappeq{\mathrel{\rlap{\raise.5ex\hbox{$<$}} {\lower.5ex\hbox{$\sim$}}}}
\def\gappeq{\mathrel{\rlap {\raise.5ex\hbox{$>$}} {\lower.5ex\hbox{$\sim$}}}}
\def\lappeq{\mathrel{\rlap{\raise.5ex\hbox{$<$}} {\lower.5ex\hbox{$\sim$}}}}

\def\GeV{{\rm GeV}}

\begin{document}

\markboth{G. Altarelli}
{Neutrino Mass and Mixing}

%
%

\title{STATUS OF NEUTRINO MASS AND MIXING
}

\author{GUIDO ALTARELLI}

\address{Dipartimento di Matematica e Fisica, Universit\`a di Roma Tre 
\\ 
INFN, Sezione di Roma Tre, \\
Via della Vasca Navale 84, I-00146 Rome, Italy
\\
and\\
CERN, Department of Physics, Theory Unit
\\ 
CH-1211 Geneva 23, Switzerland
\\
guido.altarelli@cern.ch}

\maketitle


\begin{abstract}
In the last two decades experiments have established the existence of 
neutrino oscillations and most of the related parameters have by now been measured with reasonable accuracy.  These results have accomplished a major progress  for particle physics and cosmology. At present neutrino physics is a most vital domain of particle physics and cosmology and the existing open questions are of
crucial importance. We review the present status of the subject, the main lessons that we have learnt so far and discuss the great challenges that remain in this field.
\ccode{$~~~~~~~~~~~~~~~~~~~~~~~~~~~~~~~~~~~~~~~~{\rm RM3-TH/14-6;~~~~~~CERN-PH-TH/2014-066}$}
\end{abstract}


\section{Introduction}
\label{sect:1}

The main facts on $\nu$ mass and mixing \cite{revs} are that
$\nu$'s are not all massless but their masses are very small; probably their masses are small because $\nu$'s are Majorana fermions with masses inversely proportional to the large scale M of interactions that violate lepton number (L) conservation. From the see-saw formula \cite{seesaw} together with the observed atmospheric oscillation frequency and a Dirac mass $m_D$ of the order of the Higgs vacuum expectation value (VEV), it follows that the Majorana mass scale $M \sim m_{\nu R}$ is empirically close to $10^{14}-10^{15}$ GeV $\sim M_{GUT}$, so that $\nu$ masses can well fit in the Grand Unification Theory (GUT) picture.  Decays of heavy $\nu_R$ with CP and  L violation can produce a sizable B-L asymmetry that survives instanton effects at the electroweak scale thus explaining baryogenesis as arising from leptogenesis. There is still no direct proof that neutrinos are Majorana fermions: detecting neutrino-less double beta decay ($0 \nu \beta \beta $) would prove that $\nu$'s are Majorana particles and that L is not conserved. It also appears that the active $\nu$'s are not a significant component of Dark Matter in the Universe.

On the experimental side the main recent developments on neutrino mixing \cite{revs} were the results on $\theta_{13}$ \cite{T2K,MINOS,DOUBLE-CHOOZ,RENO,DAYA-BAY} from T2K, MINOS, DOUBLE-CHOOZ, RENO  and DAYA-BAY. These experiments are in good agreement among them and the most precise is DAYA-BAY with the result \cite{DAYA-BAY} $\sin^22\theta_{13}=0.090^{+0.008}_{-0.009}$  (equivalent to $\sin^2\theta_{13}\sim 0.023\pm0.002$ or $\theta_{13}\sim(8.7\pm0.6)^o~\sim~\theta_C/\sqrt{2}$, with $\theta_C$ the Cabibbo angle).  A summary of recent global fits to the data on oscillation parameters is presented in Table \ref{ta1} \cite{Fogli:2013,GonzalezGarcia:2012} see also Ref. \citen{Tortola:2012te}. The combined value of $\sin^2\theta_{13}$ is by now about 10 $\sigma$ away from zero and the central value is rather large, close to the previous upper bound. In turn a sizable $\theta_{13}$ allows to extract an estimate of  $\theta_{23}$ from accelerator data like T2K and MINOS. There are by now solid indications of a deviation of $\theta_{23}$ from the maximal value, probably in the first octant  and, in addition, some hints of sensitivity to $\cos{\delta_{CP}}$ are starting to appear in the data \cite{Fogli:2013}.  These fits were made assuming three neutrino species but a hot issue is the possible existence of sterile neutrinos (for a review, see Ref. \citen{sterile}) that will be discussed in Sect. \ref{sect:7}.

\begin{table}[ph]
\tbl{Fits to neutrino oscillation data. For $\sin^2\theta_{23}$ from Ref. \citen{GonzalezGarcia:2012} only the absolute minimum in the first octant is shown.}
{\begin{tabular}{@{}ccc@{}} \toprule
Quantity  & Ref. \citen{Fogli:2013} & Ref. \citen{GonzalezGarcia:2012}\\
\hline
  $\Delta m^2_{sun}~(10^{-5}~{\rm eV}^2)$ &$7.54^{+0.26}_{-0.22}$ & $7.50\pm0.185$  \\
  $\Delta m^2_{atm}~(10^{-3}~{\rm eV}^2)$ &$2.43^{+0.06}_{-0.10}$ & $2.47^{+0.069}_{-0.067}$  \\
  $\sin^2\theta_{12}$ &$0.307^{+0.018}_{-0.016}$ & $0.30\pm0.013$ \\
  $\sin^2\theta_{23}$ &$0.386^{+0.024}_{-0.021}$ &  $0.41^{+0.037}_{-0.025}$ \\
  $\sin^2\theta_{13}$ &$0.0241\pm0.025$ &$0.023\pm0.0023$  \\
  \hline
\end{tabular} \label{ta1}}
\end{table}

\section{Neutrino Masses and Lepton Number Non-conservation}
\label{sect:2}

Neutrino oscillations imply non vanishing neutrino masses which in turn demand either the existence of right-handed (RH) neutrinos
(Dirac masses) or lepton number L non conservation (Majorana masses) or both. Given that neutrino masses are extremely
small, it is really difficult from the theory point of view to avoid the conclusion that L conservation must be violated.
In fact, if lepton number is not conserved the smallness of neutrino masses can be explained as inversely proportional
to the very large scale where L conservation is violated, of order $M_{GUT}$ or even $M_{Pl}$.

If L conservation is violated neutrinos are naturally Majorana fermions. For a Majorana neutrino each mass eigenstate with given helicity coincides with its own antiparticle with the same helicity. As well known, for a charged massive fermion there are four states differing by their charge and helicity (the four components of a Dirac spinor) as required by Lorentz and CPT invariance. For a massive Majorana neutrino, neutrinos and antineutrinos can be identified and only two components are needed to satisfy the Lorentz and CPT invariance constraints. Neutrinos can be Majorana fermions because, among the fundamental fermions (i.e. quarks and leptons),  they are the only electrically neutral ones. If, and only if, the lepton number L is not conserved, i.e. it is not a good quantum number, then neutrinos and antineutrinos can be identified. For Majorana neutrinos both Dirac mass terms, that conserve L ($\nu \rightarrow \nu$), and Majorana mass terms, that violate L conservation by two units ($\nu \rightarrow \bar{\nu}$), are in principle possible. Of course the restrictions from gauge invariance must be respected. So, for neutrinos the Dirac mass terms ($\bar{\nu}_R\nu_L$ +h.c.) arise from the couplings with the Higgs  field, as for all quarks and leptons. For Majorana masses, a $\nu_L^T \nu_L$ mass term has weak isospin 1 and needs two Higgs fields to make an invariant. On the contrary  a $\nu_R^T \nu_R$ mass term is a gauge singlet and needs no Higgs. As a consequence, the RH neutrino Majorana mass $M_R$ is not bound to be of the order of the electroweak symmetry breaking (induced by the Higgs VEV) and can be very large (see below).

Some notation: the charge conjugated of $\nu$ is $\nu^c$, given by $\nu^c = C(\bar{\nu})^T$, where $C=i\gamma_2 \gamma_0$ is the charge conjugation matrix acting on the spinor indices. In particular $(\nu^c)_L = C(\bar{\nu_R})^T$, so that, instead of using $\nu_L$ and $\nu_R$, we can refer to $\nu_L$ and $(\nu^c)_L $, or simply $\nu$ and $\nu^c$. 

Once we accept L non-conservation we obtain an elegant explanation for the smallness of neutrino masses. If L is not
conserved, even in the absence of heavy RH neutrinos, Majorana masses for neutrinos can be generated by dimension five
operators of the form \cite{weinberg}
\beq 
O_5=\frac{(H l)^T_i \lambda_{ij} (H l)_j}{\Lambda}~~~,
\label{O5}
\eeq  
with $H$ being the ordinary Higgs doublet, $l_i$ the SU(2) left-handed (LH) lepton doublets, $\lambda$ a matrix in  flavor space,
$\Lambda$ a large scale of mass, possibly of order $M_{GUT}$ or $M_{Pl}$ and a charge conjugation matrix $C$
between the lepton fields is understood. 
Neutrino masses generated by $O_5$ are of the order
$m_{\nu}\approx v^2/\Lambda$ for $\lambda_{ij}\approx {\rm O}(1)$, where $v\sim {\rm O}(100~\GeV)$ is the VEV of the ordinary Higgs. 

We consider that the existence of RH neutrinos $\nu^c$ is quite plausible also because most GUT groups larger than SU(5) require
them. In particular the fact that $\nu^c$ completes the representation 16 of SO(10): 16=$\bar 5$+10+1, so that all
fermions of each family are contained in a single representation of the unifying group, is too impressive not to be
significant. At least as a classification group SO(10) must be of some relevance in a more fundamental layer of the theory! Thus, in the following we assume  both that
$\nu^c$ exist and that L is not conserved. With these assumptions the see-saw mechanism \cite{seesaw} is possible.   We recall, also to fix notations, that in its simplest form it arises as follows. Consider the SU(3) $\times$ SU(2) $\times$ U(1)
invariant Lagrangian giving rise to Dirac and $\nu^c$ Majorana masses (for the time being we consider the $\nu$
(versus $\nu^c$) Majorana  mass terms as comparatively negligible):
\beq 
{\cal L}=-{\nu^c}^T y_\nu (H l)+\frac{1}{2}
{\nu^c}^T M \nu^c +~h.c.
\label{lag}
\eeq  
The Dirac mass matrix $m_D\equiv y_\nu v/\sqrt{2}$, originating from electroweak symmetry breaking,  is, in general,
non-hermitian and non-symmetric, while the Majorana mass matrix $M$ is symmetric,
$M=M^T$. We expect the eigenvalues of $M$ to be of order $M_{GUT}$ or more because $\nu^c$ Majorana masses are
SU(3)$\times$ SU(2)$\times$ U(1) invariant, hence unprotected and naturally of the order of the cutoff of the low-energy
theory.  Since all $\nu^c$ are very heavy we can integrate them away and the resulting neutrino mass matrix reads:
\beq  m_{\nu}=-m_D^T M^{-1}m_D~~~.
\eeq  

This is the well known see-saw mechanism result \cite{seesaw}: the light neutrino masses are quadratic in the Dirac
masses and inversely proportional to the large Majorana mass.  If some $\nu^c$ are massless or light they would not be
integrated away but simply added to the light neutrinos.   Note that for
$m_{\nu}\approx \sqrt{\Delta m^2_{atm}}\approx 0.05$ eV (see Table \ref{ta1}) and 
$m_{\nu}\approx m_D^2/M$ with $m_D\approx v
\approx 200~GeV$ we find $M\approx 10^{15}~GeV$ which indeed is an impressive indication for
$M_{GUT}$.

If additional contributions to $O_5$, eq. (\ref{O5}), are comparatively non-negligible, they should
simply be added.  For instance in SO(10) or in left-right extensions of the SM, an SU(2)$_L$ triplet can couple to two 
lepton doublets and to two Higgs and may induce a sizable contribution to neutrino masses. At the level of the 
low-energy effective theory, such term is still described by the operator $O_5$ of eq. (\ref{O5}),
obtained by integrating out the heavy SU(2)$_L$ triplet. This contribution is called type II
to be distinguished from that obtained by the exchange of RH neutrinos (type I). One can also have the exchange of a fermionic SU(2)$_L$ triplet coupled to a lepton doublet and a Higgs (type III).
After elimination of the heavy fields, at the level of the effective low-energy theory, the
three types of see-saw terms are equivalent. In particular they have identical transformation properties under a chiral change of
basis in flavor space. The difference is, however, that in type I see-saw mechanism, the Dirac matrix
$m_D$ is presumably related to ordinary fermion masses because they are both generated by the Higgs mechanism and both
must obey GUT-induced constraints. Thus more constraints are implied if one assumes the see-saw mechanism in its simplest type I version.

\section{Basic Formulae for Three-Neutrino Mixing}
\label{sect:3}

In this section we assume that there are only two distinct
neutrino oscillation frequencies, the atmospheric and the solar frequencies. These two can be reproduced with the known
three light neutrino species (with no need of sterile neutrinos). 

Neutrino oscillations are due to a misalignment between the flavor basis, $\nu'\equiv(\nu_e,\nu_{\mu},\nu_{\tau})$, where
$\nu_e$ is the partner of the mass and flavor eigenstate $e^-$ in a left-handed (LH) weak isospin SU(2) doublet (similarly
for 
$\nu_{\mu}$ and $\nu_{\tau}$) and the mass eigenstates $\nu\equiv(\nu_1, \nu_2,\nu_3)$ \cite{pon,lee}: 
\beq
\nu' =U \nu~~~,
\label{U}
\eeq  where $U$ is the unitary 3 by 3 mixing matrix. Given the definition of $U$ and the transformation properties of the
effective light neutrino mass matrix $m_{\nu}$ in eq. (\ref{O5}):
\bea 
\label{tr} {\nu'}^T m_{\nu} \nu'&= &\nu^T U^T m_\nu U \nu\\ \nonumber  U^T m_{\nu} U& = &{\rm
Diag}\left(m_1,m_2,m_3\right)\equiv m_{diag}~~~,
\eea  we obtain the general form of $m_{\nu}$ (i.e. of the light $\nu$ mass matrix in the basis where the charged lepton
mass is a diagonal matrix):
\beq  m_{\nu}=U^* m_{diag} U^\dagger~~~.
\label{gen}
\eeq  The matrix $U$ can be parametrized in terms of three mixing angles $\theta_{12}$,
$\theta_{23}$ and $\theta_{13}$ ($0\le\theta_{ij}\le \pi/2$)  and one phase $\varphi$ ($0\le\varphi\le 2\pi$) \cite{cab},
exactly as for the quark mixing matrix $V_{CKM}$. The following definition of mixing angles can be adopted:
\beq  U~=~ 
\left(\begin{matrix}1&0&0 \cr 0&c_{23}&s_{23}\cr0&-s_{23}&c_{23}\end{matrix}     
\right)
\left(\begin{matrix}c_{13}&0&s_{13}e^{i\varphi} \cr 0&1&0\cr -s_{13}e^{-i\varphi}&0&c_{13} \end{matrix} 
\right)
\left(\begin{matrix}c_{12}&s_{12}&0 \cr -s_{12}&c_{12}&0\cr 0&0&1    \end{matrix}
\right)
\label{ufi}
\eeq  where $s_{ij}\equiv \sin\theta_{ij}$, $c_{ij}\equiv \cos\theta_{ij}$.  In addition, if $\nu$ are Majorana particles,
we have the relative phases among the Majorana masses
$m_1$, $m_2$ and $m_3$.  If we choose $m_3$ real and positive, these phases are carried by $m_{1,2}\equiv\vert m_{1,2}
\vert e^{i\phi_{1,2}}$
\cite{frsm}.  Thus, in general, 9 parameters are added to the SM when non-vanishing neutrino masses are included: 3
eigenvalues, 3 mixing angles and 3 CP  violating phases.

In our notation the two frequencies, $\Delta m^2_{I}/4E$ $(I$=sun,atm), are parametrized in terms of the $\nu$ mass
eigenvalues by 
\beq
\Delta m^2_{sun}\equiv \vert\Delta m^2_{12}\vert ,~~~~~~~
\Delta m^2_{atm}\equiv \vert\Delta m^2_{23}\vert~~~.
\label{fre}
\eeq   where $\Delta m^2_{12}=\vert m_2\vert^2-\vert m_1\vert^2 > 0$ (positive by the definition of $m_{1,2}$) and $\Delta m^2_{23}= m_3^2-\vert m_2\vert ^2$. The
numbering 1,2,3 corresponds to our definition of the frequencies and in principle may not coincide with the ordering from
the lightest to the heaviest state. In fact, the sign of $\Delta m^2_{23}$ is not known [a positive (negative) sign corresponds to normal (inverse) hierarchy].
The determination of the hierarchy pattern together with the measurement of the CP violating phase $\varphi$ are among the main experimental challenges for future accelerators.

Oscillation experiments do not provide information about the absolute neutrino mass scale. Limits on that are obtained \cite{revs} from the endpoint of the tritium beta decay spectrum, from cosmology and from neutrinoless double beta decay ($0\nu \beta \beta$). From tritium we have an absolute upper limit of
2.2 eV (at 95\% C.L.) \cite{pdg} on the antineutrino mass eigenvalues involved in beta decay, which, combined with the observed oscillation
frequencies under the assumption of three CPT-invariant light neutrinos, also amounts to an upper bound on the masses of
the other active neutrinos. The near-future of the tritium measurement is the KATRIN experiment whose goal is to improve the present limit by about an order of magnitude \cite{katrin}. Complementary information on the sum of neutrino masses is also provided by cosmology. For the sum of all (quasi) stable (thermalized) neutrino masses the Planck experiment, also using the WMAP-9 and BAO data, finds the limit $\sum m_\nu \leq 0.23$ at 95$\%$ c.l. \cite{Pl}. The discovery of $0\nu \beta \beta$ decay would be very important, as discussed in the next section, and would also provide direct information on the absolute
scale of neutrino masses.

\section{Importance of Neutrino-less Double Beta Decay}
\label{sect:4}

The detection of neutrino-less double beta decay \cite{zuber} would provide direct evidence of $L$ non conservation and of the Majorana nature of neutrinos. It would also offer a way to possibly disentangle the 3 cases of degenerate, normal or inverse hierachy neutrino spectrum.  The quantity which is bound by experiments on $0\nu \beta \beta$
is the 11 entry of the
$\nu$ mass matrix, which in general, from $m_{\nu}=U^* m_{diag} U^\dagger$, is given by :
\bea 
\vert m_{ee}\vert~=\vert(1-s^2_{13})~(m_1 c^2_{12}~+~m_2 s^2_{12})+m_3 e^{2 i\phi} s^2_{13}\vert
\label{3nu1gen}
\eea
where $m_{1,2}$ are complex masses (including Majorana phases) while $m_3$ can be taken as real and positive and $\phi$ is the $U$ phase measurable from CP violation in oscillation experiments. Starting from this general formula it is simple to
derive the bounds for degenerate, inverse hierarchy or normal hierarchy mass patterns. 

At present the best limits from the searches with Ge lead to $\vert m_{ee}\vert~\sim ~(0.25-0.98)$ eV (GERDA \cite{gerda}+HM \cite{HM}+IGEX \cite{igex}) and with Xe to $\vert m_{ee}\vert~\sim ~(0.12-0.25)$ eV (EXO \cite{exo}+Kamland Zen \cite{kamzen}), where ambiguities on the nuclear matrix elements lead to the ranges shown.
In the next few years, experiments (CUORE, GERDA II, SNO+....) will reach a larger sensitivity on $0\nu \beta \beta$ by about an order of magnitude. Assuming the standard mechanism through mediation of a light massive Majorana neutrino, if these experiments will observe a signal this would indicate that the inverse hierarchy is realized, if not, then the normal hierarchy case still would remain a possibility. 

\section{Baryogenesis via Leptogenesis from Heavy $\nu_R$ Decay}
\label{sect:5}

In the Universe we observe an apparent excess of baryons over antibaryons. It is appealing that one can explain the
observed baryon asymmetry by dynamical evolution (baryogenesis) starting from an initial state of the Universe with zero
baryon number.  For baryogenesis one needs the three famous Sakharov conditions: B violation, CP violation and no thermal
equilibrium. In the history of the Universe these necessary requirements have probably occurred at different epochs. Note
however that the asymmetry generated during one such epoch could be erased in following epochs if not protected by some dynamical
reason. In principle these conditions could be fulfilled in the SM at the electroweak phase transition. In fact, when kT is of the order of a few TeV, B conservation is violated by
instantons (but B-L is conserved), CP symmetry is violated by the CKM phase and
sufficiently marked out-of- equilibrium conditions could be realized during the electroweak phase transition. So the
conditions for baryogenesis  at the weak scale in the SM superficially appear to be present. However, a more quantitative
analysis
\cite{tro} shows that baryogenesis is not possible in the SM because there is not enough CP violation and the phase
transition is not sufficiently strong first order, because the Higgs mass is too heavy. In SUSY extensions of the SM, in particular in the MSSM,
there are additional sources of CP violation but also this possibility has by now become at best marginal after the results from LEP2 and the LHC.

If baryogenesis at the weak scale is excluded by the data still it can occur at or just below the GUT scale, after inflation.
But only that part with
$|{\rm B}-{\rm L}|>0$ would survive and not be erased at the weak scale by instanton effects. Thus baryogenesis at
$kT\sim 10^{10}-10^{15}~{\rm GeV}$ needs B-L violation and this is also needed to allow $m_\nu$ if neutrinos are Majorana particles.
The two effects could be related if baryogenesis arises from leptogenesis then converted into baryogenesis by instantons
\cite{fuku,buch}. The decays of heavy Majorana neutrinos (the heavy eigenstates of the see-saw mechanism) happen with non conservation of lepton number L, hence also of B-L and can well involve a sufficient amount of CP violation. Recent results on neutrino masses are compatible with this elegant possibility. Thus the case
of baryogenesis through leptogenesis has been boosted by the recent results on neutrinos.

\section{A Drastic Conjecture: the $\nu$MSM}
\label{sect:6}

The most direct version of the see-saw mechanism with heavy $\nu_R$ matches well with GUT's (e.g. the heavy Majorana mass is given by $M_\nu \sim M_{GUT}$, we observe a complete 16 of SO(10) for each generation...). As well known, for naturalness, one would expect a completion of the SM near the EW scale in order to understand the big gap between $m_H$ and $M_{GUT}$. The most attractive and well studied example of this sort of enlargement is that of supersymmetry (SUSY). Within SUSY one also has excellent candidates for Dark Matter and the GUT picture is improved by a precise gauge coupling unification, compatibility of the predicted proton lifetime with existing bounds etc. However no SUSY nor any other form of new physics has been found at the LHC or elsewhere and, as a consequence, our concept of naturalness has so far failed as a heuristic principle \cite{sosimple}. While SUSY remains an attractive possibility with perhaps a still acceptable degree of fine-tuning, it is true that models where the fine-tuning problem is disregarded or reconsidered have been revived.

It is important to note that although the hierarchy problem is directly related to the quadratic divergences in the scalar sector of the SM, actually the problem can be formulated without any reference to divergences or to a cut-off, directly in terms of renormalized quantities. After renormalization the hierarchy problem is manifested by the quadratic sensitivity of  the scalar sector mass scale $\mu^2$ to the physics at large energy scales. If there is a threshold at large energy, where some particles of mass $M$ coupled to the Higgs sector can be produced and contribute in loops, then the renormalized running mass $\mu$ would evolve slowly (i.e. logarithmically according to the relevant beta functions \cite{strubeta}), up to $M$ and there, as an effect of the matching conditions at the threshold, rapidly jump to become of order $M$ (see, for example, \cite{barThr}).  In the presence of a threshold at $M$ one needs a fine tuning of order $\mu^2/M^2$ in order to reproduce the observed value of the running mass at low energy. Note that heavy RH neutrinos, which are coupled to the Higgs through the Dirac Yukawa coupling, would contribute in the loop and, in the absence of SUSY,  become unnatural at $M \gappeq 10^7 - 10^8$ GeV \cite{nuunnat}. Also, in the pure Standard Model heavy $\nu_R$ tend to destabilize the vacuum and make it unstable for $M \gappeq 10^{14}$ GeV \cite{nudest}. Thus for naturalness either new thresholds appear but there is a mechanism for the cancellation of the sensitivity (e.g. a symmetry like SUSY) or they would better not appear at all, except that certainly there is the Planck mass, connected to the onsetting of quantum gravity, that sets an unavoidable threshold. A possible point of view is that there are no new thresholds up to $M_{Planck}$ (at the price of giving up GUTs, among other things) but, miraculously, there is a hidden mechanism in quantum gravity that solves the fine tuning problem related to the Planck mass \cite{shapo,gian}. For this one would need to solve all phenomenological problems, like Dark Matter, baryogenesis and so on, with physics below the EW scale.  This point of view is extreme but allegedly not yet ruled out. Possible ways to realize this program are discussed in Ref. \citen{shapo}: one has to introduce three RH neutrinos, $N_1$, $N_2$ and $N_3$ which are now light: for $N_1$ we need $m_1$ ~few keV, while $m_{2,3}$ ~ few GeV but with a few eV splitting. With this rather ad hoc spectrum $N_1$ can explain Dark Matter and $N_{2,3}$ baryogenesis. The active neutrino masses are obtained from the see-saw mechanism, but with very small Dirac Yukawa couplings. Then the data on neutrino oscillations can be reproduced. The RH $N_i$ can give rise to observable consequences (and in fact only a limited domain of the parameter space is still allowed). In fact $N_1$ could decay as $N_1 \rightarrow \nu + \gamma$ producing a line in X-ray spectra at $E_\gamma \sim m_1/2$. It is interesting that a candidate line with $E_\gamma \sim 3.5$ keV has been identified in the data of the XMM-Newton X-ray observatory on the spectra from galaxies or galaxy clusters \cite{line}.  As for $N_{2,3}$ they could be looked for in charm meson decays if sufficiently light. A Letter of Intent for a dedicated experiment at the CERN SpS has been presented to search for these particles \cite{loi}. 

\section{Oscillations with Sterile Neutrinos}
\label{sect:7}

A number of hints have been recently collected in neutrino oscillation experiments for the existence of sterile neutrinos, that is neutrinos with no weak interactions  (for a review see Ref. \citen{white}). They do not make yet an evidence but certainly pose an experimental problem that needs clarification (see, for example, Ref. \citen{rub}). 

The MiniBooNE experiment published  \cite{Mboo:2012} a combined analysis of  $\nu_e$ appearance in a $\nu_\mu$ beam together with  $\bar{\nu}_e$ appearance in a $\bar{\nu}_\mu$ beam. They observe
an excess of events from neutrinos over expected backgrounds in the low energy  region (below ~500 MeV) of the
spectrum. In the most recent data
the shapes of the neutrino and anti-neutrino spectra appear to be consistent with each
other, showing excess events below ~500 MeV and data consistent with background in
the high energy region. The allowed region from MiniBooNE anti-neutrino data has some overlap with the parameter region preferred by LSND \cite{LSND}. Recently the ICARUS \cite{Antonello:2012} and OPERA \cite{agafo} experiments at Gran Sasso have published the results of their searches for electrons produced by the CERN neutrino beam. No excess over the background was observed. As a consequence a large portion of the region allowed by LSND, MiniBooNE. KARMEN \cite{karmen}... is now excluded.

Then there are $\bar \nu_e$ disappearance experiments: in particular, the reactor and the gallium anomalies. A reevaluation of the reactor flux \cite{Mention:2011} produced an apparent gap between the theoretical expectations and the data taken at small distances from  reactors ($\le$ 100 m). A different analysis confirmed the normalization shift \cite{Huber:2011}. Similarly the Gallium anomaly \cite{Giunti:2010} depends on the assumed cross-section which could be questioned. 

These data hint at one or more sterile neutrinos with mass around $\sim$ 1 eV which would
represent a major discovery in particle physics. Cosmological data would certainly allow for one sterile neutrino while more than one are disfavored by the stringent bounds arising form nucleosynthesis (assuming fully thermalized sterile neutrinos)  \cite{GiuJou}. Actually the recently published Planck data \cite{Pl} on the cosmic microwave background (CMB) are completely consistent with no sterile neutrinos (they quote for the total number of neutrinos $N_{eff}= 3.31\pm0.53$). The absence of a positive signal in $\nu_\mu$ disappearance in accelerator experiments (MINOS \cite{minosste},  MiniBooNE-SciBooNE \cite{minisci}) creates a tension with LSND (if no CP viol.). For example, in 3+1 models there is a tension \cite{kopp} between appearance (LSND, MiniBooNe.....) and disappearance (MINOS...) . However, the 3+1 fit is much improved if 
the low energy MiniBooNe data are not included \cite{laveder}. In 3+1 models the short baseline reactor data and the gallium anomaly are not in tension with the other measurements. Fits with 2 sterile neutrinos do not solve all the tensions \cite{kopp,conrad}. In general in all fits the resulting sterile neutrino masses are a bit too large when compared with the cosmological bounds on the sum of neutrino masses, if the contribution of the sterile neutrinos to the effective number of relativistic degrees of freedom is close to one.

In conclusion, the situation is at present confuse but the experimental effort should be continued  because establishing the existence of sterile neutrinos would be a great discovery (an experiment to clarify the issue of sterile neutrinos is proposed on the CERN site \cite{ant}). In fact a sterile neutrino is an exotic particle not predicted by the most popular models of new physics.

As only a small leakage from active to sterile neutrinos is allowed by present neutrino oscillation data (see, for example, refs. \cite{Archidiacono:2013,Palazzo,kopp,Mirizzi} and references therein), in the following we restrict our discussion to 3-neutrino models.

\section{Models of Neutrino Mixing}
\label{sect:8}

A long list of models have been formulated
over the years to understand neutrino mixing. With time and the continuous
improvement of the data most of the models have been discarded by experiment. But the surviving models still span a wide range going from a maximum of
symmetry, e.g. with discrete non-abelian flavor groups, to the opposite extreme of
Anarchy.

The relatively large measured value of $\theta_{13}$, close in size to the Cabibbo angle, and the indication that $\theta_{23}$ is not maximal both go in the direction of models based on Anarchy \cite{Hall:1999sn,deGouvea:2003xe}, i.e. the idea that perhaps no symmetry is needed in the neutrino sector, only chance (this possibility has been recently reiterated, for example, in Ref. \citen{deGouvea:2012ac}). The appeal of Anarchy is augmented if formulated in a $SU(5) \otimes U(1)_{FN}$ context with different Froggatt-Nielsen \cite{Froggatt:1978nt} charges only for the $SU(5)$ tenplets (for example $10\sim(a,b,0)$, where $a > b > 0$ is the charge of the first generation, b of the second, zero of the third) while no charge differences appear in the $\bar 5$ (e. g. $\bar 5\sim (0,0,0)$). In fact, the observed fact that the up-quark mass hierarchies are more pronounced than for down-quark and charged leptons is in agreement with this assignment. Indeed the embedding of Anarchy in the $SU(5) \otimes U(1)_{FN}$ context allows to implement a parallel  treatment of quarks and leptons. Note that implementing Anarchy and its variants in $SO(10)$ would be difficult. In models with no see-saw, the $\bar 5$ charges completely fix the hierarchies (or Anarchy, if the case) in the neutrino mass matrix.  If RH neutrinos are added, they transform as $SU(5)$ singlets and can in principle carry independent $U(1)_{FN}$ charges, which also, in the Anarchy case, must be all equal. With RH neutrinos the see-saw mechanism can take place and the resulting phenomenology is modified.   

The $SU(5)$ generators act
ÔverticallyÕ inside one generation, whereas the $U(1)_{FN}$ charges differ ÔhorizontallyÕ from one
generation to the other. If, for a given interaction vertex, the $U(1)_{FN}$  charges do not add to zero, the
vertex is forbidden in the symmetric limit. However, the $U(1)_{FN}$ symmetry (that one can assume to be a gauge symmetry) is spontaneously broken by
the VEVs $v_f$ of a number of ÔflavonÕ fields with non-vanishing charge and GUT-scale masses. Then a forbidden coupling
is rescued but is suppressed by powers of the small parameters $\lambda = v_f/M$, with $M$ a large mass, with the exponents larger for larger charge mismatch. Thus the charges fix the powers of $\lambda$, hence the degree of suppression of all elements of mass matrices, while arbitrary coefficients $k_{ij}$ of order 1 in each entry of mass matrices are left unspecified (so that the number of order 1 parameters exceeds the number of observable quantities). A random selection of these $k_{ij}$ parameters leads to distributions of resulting values for the measurable quantities. For Anarchy the mass matrices in the neutrino sector (determined by the $\bar 5$ and $1$ charges) are totally random, while in the presence of unequal charges different entries carry different powers of the order parameter and thus suitable hierarchies are enforced for quarks and charged leptons. 

Within this framework there are many variants of models largely based on chance: fermion charges can all be nonnegative
with only negatively charged flavons, or there can be fermion charges of different signs
with either flavons of both charges or only flavons of one charge.
 In Refs.\citen{AFMM:2012,melmer}, given the new experimental results, a reappraisal of Anarchy and its variants within the $SU(5)\times U(1)_{\rm FN}$ GUT framework was made. Based on the most recent data it is argued  that the Anarchy ansatz is probably oversimplified and, in any case, not compelling. In fact, suitable differences of $U(1)_{FN}$ charges, if also introduced within pentaplets and singlets, lead, with the same number of random parameters as for Anarchy, to distributions that are in better agreement with the data. The hierarchy of quark masses and mixing and of charged lepton masses in all cases impose a hierarchy-defining parameter of the order of $\lambda_C=\sin{\theta_C}$. The weak points of Anarchy are that all mixing angles should be of the same order, so that the relative smallness of $\theta_{13}\sim o(\lambda_C)$ is not automatic. Similarly the smallness of $r=\Delta m^2_{solar}/\Delta m^2_{atm}\sim (0.175)^2$ is not easily reproduced: with no see-saw $r$ is expected of $o(1)$, while in the see-saw version of Anarchy the problem is only partially alleviated by the spreading of the neutrino mass distributions that follows from the product of three matrix factors in the see-saw formula. An advantage is already obtained if Anarchy is only restricted to the 23 sector of leptons. In this case, with or without see-saw,  $\theta_{13}$ is naturally suppressed and, with a single fine tuning one gets both $\theta_{12}$ large and $r$ small (this model was also recently rediscussed in Ref. \citen{Buchmuller:2011tm}). Actually in Ref. \citen{AFMM:2012} it was shown, for example,  that the freedom of adopting RH neutrino charges of  both signs, can be used to obtain a completely natural model where all small quantities are suppressed by the appropriate power of $\lambda_C$. In this model a lopsided Dirac mass matrix is combined with a generic Majorana matrix to produce a neutrino mass matrix where the 23 subdeterminant is suppressed and thus $r$ is naturally small and $\theta_{23}$ is large. In addition  also $\theta_{12}$ is large while $\theta_{13}$ is suppressed. We stress again that the number of random parameters is the same in all these models: one coefficient of $o(1)$ for every matrix element. But, with an appropriate choice of charges, the observed order of magnitude of all small parameters can be naturally explained and the charged fermion hierarchies and the quark mixing angles can be  accommodated.  In conclusion, models based on chance are still perfectly viable, but we consider Anarchy a particularly simple choice, perhaps oversimplified and certainly not compelling, and we have argued that, since the hierarchy of charged fermion masses needs a minimum of flavor symmetry (like $U(1)_{FN}$) then it is plausible that, to some extent, this flavor symmetry can also be effective in the neutrino sector.

Anarchy and its variants, all sharing the dominance of randomness in the lepton sector, are to be confronted with models with a richer dynamical structure, some based on continuous groups \cite{cont} but in particular those based on discrete flavor groups (for  reviews, see, for example, Refs.~\cite{Altarelli:2010gt,ishikilu,grilu}). After the measurement of a relatively large value for $\theta_{13}$ there has been an intense work to interpret the new data along different approaches and ideas.  
Examples are suitable modifications of the minimal models \cite{lessmin,Lin:2009bw} (we discuss the Lin model of Ref. \citen{Lin:2009bw} in the following), modified sequential dominance models  \cite{steve}, larger symmetries that  already at LO  lead to non vanishing $\theta_{13}$ and non maximal $\theta_{23}$ \cite{altmix}, smaller symmetries that leave more freedom \cite{lesssimm}, models where the flavor group and a generalised CP transformation are combined in a non trivial way \cite{cpfla} (other approaches to discrete symmetry and CP violation are found in Refs. \cite{othercp}).

\begin{figure}[t]
\centerline{\includegraphics[width=10 cm]{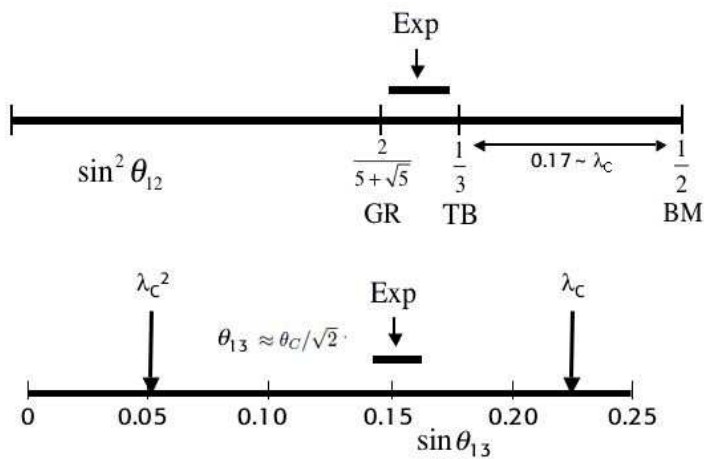}}
\caption{Top: the experimental value of $\sin^2{\theta_{12}}$ is compared with the predictions of exact Tri-Bimaximal (TB) or Golden Ratio (GR) or Bi-Maximal mixing (BM). The shift needed to bring the TB or the GR predictions to agree with the experimental value is small, numerically of order $\lambda_C^2$, while it is larger, of order $\lambda_C$ for the BM case, where $\lambda_C \equiv \sin{\theta_C}$ with $\theta_C$ being the Cabibbo angle. Bottom:  the experimental value of $\sin{\theta_{13}}$ in comparison with $\lambda_C$ or $\lambda_C^2$. \label{f1}}
\end{figure}

 Among the models with a non trivial dynamical structure those based on discrete flavor groups were motivated by the fact that the data suggest some special mixing patterns as good first approximations like Tri-Bimaximal (TB) or Golden Ratio (GR) or Bi-Maximal (BM) mixing, for example. The corresponding mixing matrices all have $\sin^2{\theta_{23}}=1/2$, $\sin^2{\theta_{13}}=0$, values that are good approximations to the data (although less so since the most recent data), and differ by the value of the solar angle $\sin^2{\theta_{12}}$ (see Fig.~\ref{f1}). The observed $\sin^2{\theta_{12}}$, the best measured mixing angle,  is very close, from below, to the so called Tri-Bimaximal (TB) value \cite{Harrison} of $\sin^2{\theta_{12}}=1/3$ \footnote{A model proposed by Fritzsch and Zing in Refs. \citen{fz} can be considered as an ancestor of  TB mixing but with $\theta_{12}$ and $\theta_{23}$ interchanged, which is not supported by the present data.}. Alternatively, it is also very close, from above, to the Golden Ratio (GR) value \cite{GR1} $\sin^2{\theta_{12}}=\frac{1}{\sqrt{5}\,\phi} = \frac{2}{5+\sqrt{5}}\sim 0.276$, where $\phi= (1+\sqrt{5})/2$ is the GR (for a different connection to the GR, see Refs.~\cite{GR2}). On a different perspective, one has also considered models with Bi-Maximal (BM) mixing, where at leading order (LO), before diagonalization of charged leptons, $\sin^2{\theta_{12}}=1/2$, i.e. it is also maximal,  and the necessary, rather large, corrective terms to $\theta_{12}$ arise from the diagonalization of the charged lepton mass matrices \cite{Altarelli:2004jb} (a long list of references can be found in Ref.~\citen{Altarelli:2010gt}). Thus, if one or the other of these coincidences is taken seriously, models where TB or GR or BM mixing is naturally predicted provide a good first approximation (but these hints cannot all be relevant and it is well possible that none is).
As the corresponding mixing matrices have the form of rotations with fixed special angles one is naturally led to discrete flavor groups. 

In the following we will mainly refer to TB or BM mixing which are the most studied first approximations to the data. A simplest discrete symmetry for TB mixing is $A_4$ while BM can be obtained from $S_4$. 
Starting with the ground breaking paper in Ref. \citen{Ma:2001},
$A_4$ models  have been widely studied  (for a recent review and a list of references, see Ref.~\citen{Altarelli:2012}). At LO the typical $A_4$ model (like, for example, the one discussed in Ref. \citen{Altarelli:2006ty}) leads to exact TB mixing.  In these models the starting LO approximation is completely fixed (no chance), but the Next to LO (NLO) corrections, which are specified by the set of flavor symmetries and the field content of the model, still introduce a number of undetermined parameters, although in general much less in number than for $U(1)_{FN}$ models. These models are therefore more predictive and in each model, one obtains relations among the departures of the three mixing angles from the LO patterns, restrictions on the CP violation phase $\delta_{CP}$, mass sum rules among the neutrino mass eigenvalues, definite ranges for the neutrinoless beta decay effective Majorana mass and so on. 
In the absence of specific dynamical tricks, in a generic model at NLO all three mixing angles receive corrections of the same order of magnitude. Since the experimentally allowed departures of
 $\theta_{12}$ from the TB value, $\sin^2{\theta_{12}}=1/3$, are small, numerically not larger than $\mathcal{O}(\lambda_C^2)$ where $\lambda_C=\sin\theta_C$, it follows that both $\theta_{13}$ and the deviation of $\theta_{23}$ from the maximal value are also expected to be typically of the same general size.  This generic prediction of a small $\theta_{13}$, numerically of  $\mathcal{O}(\lambda_C^2)$, is at best marginal after the recent measurement of $\theta_{13}$.
 
Of course, one can introduce some additional theoretical input to improve the value of $\theta_{13}$ \cite{AFMS:2012}. In the case of $A_4$, one particularly interesting example is provided by the  Lin model \cite{Lin:2009bw} (see also Ref. \citen{lessmin}), formulated before the recent $\theta_{13}$ results.  In the Lin model the $A_4$ symmetry breaking is arranged, by suitable additional $Z_n$ parities, in a way that the corrections to the charged lepton and the neutrino sectors are kept separated not only at LO but also at NLO. As a consequence, in a natural way the contribution to neutrino mixing from the diagonalization of the charged leptons can be of $\mathcal{O}(\lambda_C^2)$, while those in the neutrino sector of $\mathcal{O}(\lambda_C)$. Thus, in the Lin model the NLO corrections to the solar angle $\theta_{12}$ and to the reactor angle $\theta_{13}$ are not necessarily related. In addition, in the Lin model the largest corrections do not affect $\theta_{12}$ and satisfy the relation $\sin^2{\theta_{23}} =1/2 +1/\sqrt{2}\cos{\delta_{CP}} |\sin{\theta_{13}}|$, with $\delta_{CP}$ being the CKM-like CP violating phase of the lepton sector. Note that, for $\theta_{23}$ in the first octant, the sign of  $\cos{\delta_{CP}}$ must be negative.  

Alternatively, one can think of models where, because of a suitable symmetry,  BM mixing holds in the neutrino sector at LO and the corrective terms for $\theta_{12}$, which in this case are required to be large, arise from the diagonalization of charged lepton masses (for a list of references, see Ref.~\citen{Altarelli:2012})). These terms from the charged lepton sector, numerically required of order $\mathcal{O}(\lambda_C)$, would then generically also affect $\theta_{13}$ and the resulting angle could well be compatible with the measured value. Thus $\theta_{13}$ large is not a problem in this class of models. An explicit model of this type based on the group $S_4$ has been developed in Ref.~\citen{Altarelli:2009gn} (see also Refs.~\citen{bms4}). In analogy with the CKM mixing of quarks one assumes that the 12 entry of the charged lepton diagonalization matrix is dominant and of order $\theta_C$. An important feature of this particular model is that only $\theta_{12}$ and $\theta_{13}$ are corrected by terms of $\mathcal{O}(\lambda_C)$ while $\theta_{23}$ is unchanged at this order.  Note however that, in a supersymmetric context (for a recent general analysis of LFV effects in the context of flavor models, see Ref.~\citen{Calibbi:2012at}), the  present bounds on lepton flavor violating (LFV) reactions pose severe constraints on the parameter space of the models. In particular, we refer to the recent improved MEG result \cite{meg} on the $\mu \rightarrow e \gamma$ branching ratio, $Br(\mu \rightarrow e \gamma) \leq 5.7\times10^{-13}$ at $90\%$ C.L. and to other similar processes like $\tau \rightarrow (e~\rm{or}~ \mu)  \gamma$. Particularly constrained are the models with  relatively large corrections from the off-diagonal terms of the charged lepton mass matrix, like the models with BM mixing at LO \cite{AFMS:2012}. A way out, indicated by the failure to discover SUSY at the LHC,  is to push the s-partners at large enough masses but then a supersymmetric explanation of the muon (g-2) anomaly becomes less plausible \cite{amu,HMpdg}.

In conclusion, one could have imagined that neutrinos would bring a decisive
boost towards the formulation of a comprehensive understanding of fermion masses
and mixings. In reality it is frustrating that no real illumination was sparked on the
problem of flavor. We can reproduce in many different ways the observations, in a
wide range that goes from anarchy to discrete flavor symmetries but we have not
yet been able to single out a unique and convincing baseline for the understanding of
fermion masses and mixings. In spite of many interesting ideas and the formulation
of many elegant models the mysteries of the flavor structure of the three generations
of fermions have not been much unveiled.

\section{Conclusion}
\label{sect:9}

Neutrino physics deals with fundamental issues still of great importance.
Our knowledge of neutrino physics has been much
advanced in the last 15 years and it is still vigorously studied and progress is continuously made,
but many crucial problems are still open. Together with LHC physics the study of neutrino and flavor processes maintains a central role in fundamental physics.

\section*{Acknowledgments}
I thank the Organizers, in particular Harald Fritzsch and Kok Khoo Phua, for their kind invitation and hospitality in Singapore.
I recognize that this work has been partly supported by the COFIN program (PRIN 2008) and by the European Commission, under the networks ÒLHCPHENONETÓ and ÒInvisiblesÓ

\end{document}